\documentclass{pasj00}
\draft
\begin{document}
\SetRunningHead{N. Isobe et al.}
{Ultraluminous X-ray source, NGC 2403 Source 3}      
\Received{yyyy/mm/dd}
\Accepted{yyyy/mm/dd}

\title{Spectral transitions of an ultraluminous X-ray source, 
NGC 2403 Source 3}

\author{%
Naoki       \textsc{Isobe}      \altaffilmark{1},
Kazuo       \textsc{Makishima}  \altaffilmark{1,2},
Hiromitsu   \textsc{Takahashi}  \altaffilmark{3},
Tsunefumi   \textsc{Mizuno}     \altaffilmark{3},\\
Ryouhei     \textsc{Miyawaki}   \altaffilmark{2},
Poshak      \textsc{Gandhi}     \altaffilmark{1},
Madoka      \textsc{Kawaharada} \altaffilmark{1},
Atsushi     \textsc{Senda}      \altaffilmark{1},\\
Tessei      \textsc{Yoshida}    \altaffilmark{4,5},
Aya         \textsc{Kubota}     \altaffilmark{6},
Hiroshi     \textsc{Kobori}     \altaffilmark{6},
}
\altaffiltext{1}{Cosmic Radiation Laboratory,
   the Institute of Physical and Chemical Research, \\ 
   2-1 Hirosawa, Wako, Saitama, 351-0198, Japan}
\email{isobe@crab.riken.jp}
\altaffiltext{2}{Department of Physics, University of Tokyo,
   7-3-1 Hongo, Bunkyo-ku, Tokyo, Japan}
\altaffiltext{3}{Department of Physics, Hiroshima University, \\
   1-3-1 Kagamiyama, Higashi-Hiroshima, Hiroshima 739-8526, Japan}
\altaffiltext{4}{Institute of Space and Astronautical Science,
Japan Aerospace Exploration Agency (ISAS/JAXA),\\
3-1-1 Yoshinodai, Sagamihara-shi, Kanagawa 229-8510, Japan}
\altaffiltext{5}{Department of Physics, Tokyo University of Science,\\ 
1-3 Kagurazaka, Shinjuku-ku, Tokyo 162-8601, Japan}
\altaffiltext{6}{Department of Electronic Information Systems,
   Shibaura Institute of Technology, \\
   307 Fukasakum Minuma-ku, Saitama-shi, Saitama, 337-8570, Japan}

\maketitle
\begin{abstract}
Suzaku observation of an ultraluminous X-ray source,                  
NGC 2403 Source 3, performed on 2006 March 16--17, is reported.       
The Suzaku XIS spectrum of Source 3 was described                     
with a multi-color black-body-like emission                           
from an optically thick accretion disk.                               
The innermost temperature and radius of the accretion disk            
was measured to be $T_{\rm in} = 1.08_{-0.03}^{+0.02} $ keV and          
$R_{\rm in} = 122.1_{-6.8}^{+7.7}~\alpha^{1/2}$ km, respectively,        
where $\alpha = (\cos 60^\circ /\cos i)$                              
with $i$ being the disk inclination.                                  
The bolometric luminosity of the source was estimated to be           
$L_{\rm bol} = 1.82 \times 10^{39} \alpha $ ergs s$^{-1}$.              
Archival Chandra and XMM-Newton data of the source                   
were analyzed for long-term spectral variations.                      
In almost all observations,                                           
the source showed multi-color black-body-like X-ray spectra           
with parameters similar to those in the Suzaku observation.           
In only one Chandra observation, however,                             
Source 3 was found to exhibit a power-law-like spectrum,              
with a photon index of $\Gamma = 2.37 \pm 0.08$,                      
when it was fainter by about $\sim 15 \%$ than                        
in the Suzaku observation.                                            
The spectral behavior is naturally explained                          
in terms of a transition between the slim disk state and              
the ``very high'' states, both found in Galactic black hole binaries  
when their luminosity approach the Eddington limit.                   
These results are utilized to argue that ultraluminous X-ray sources   
generally have significantly higher black-hole masses                 
than ordinary stellar-mass black holes.                               
\end{abstract}

\section{Introduction} 
\label{sec:intro}
Since the era of the Einstein observatory \citep{ULX_Einstein},
luminous point-like X-ray sources,
of which X-ray luminosities exceed a few times $10^{39}$ ergs s$^{-1}$, 
were frequently found at off-center regions of nearby normal galaxies.  
These X-ray sources are called ultraluminous X-ray Sources 
(ULXs; \cite{ULX_asca1}). 
The nature of ULXs has remained 
one of the important unresolved issues in X-ray astrophysics,
for nearly three decades.

Both observational 
(e.g., \cite{ULX_asca2,XTEJ1550,M81X9,NGC1313_Suzaku,SuzakuJ1305})
and theoretical (e.g., \cite{slimdisk}) studies suggest that
the X-ray properties of ULXs resemble 
those of Galactic black hole binaries (BHBs) in high accretion rates. 
However, we currently have two distinct interpretations of ULXs; 
intermediate mass BHBs with a mass of $M \gg 10 M_\odot$ 
(where $M_\odot$ is the solar mass) radiating 
at sub- or trans-Eddington luminosities,
and 
stellar-mass BHBs shining at highly super-Eddington luminosities. 
We urgently need to distinguish the two alternative. 

X-ray spectra of well-studied ULXs are approximated 
by  either a multi-color disk (MCD; \cite{MCD1,MCD2}) model 
or a power-law (PL) one (or their combination).
Furthermore, several ULXs are reported to show a spectral transition 
between the MCD-like and PL-like states (e.g., \cite{IC342_transition}). 
These properties could be explained, in the simplest manner, 
by regarding the MCD-like and PL-like states 
as analogous to the classical high/soft and low/hard states
of Galactic BHBs, respectively. 
However, this naive interpretation has encountered several difficulties
(e.g., \cite{ULX_asca1,ULX_asca2,IC342_VHS}).

Recent observational studies (e.g, \cite{M81X9})
revealed that the X-ray spectra of the MCD-like ULX 
requires a temperature profile in the accretion disk 
that is flatter than those in the standard accretion disk 
\citep{standard_disk} assumed in the MCD model. 
The flat temperature profile is theoretically predicted 
by a concept of slim accretion disk \citep{slimdisk},  
in which an optically-thick advection
and/or photon trapping (e.g., \cite{photon-trapping})
become important due to high accretion rates.
The slim disk interpretation can solve several basic problems
with the MCD model, including too-high disk temperatures  
and variable innermost disk radii \citep{ULX_asca2,slimdisk} .
Therefore, the MCD-like ULX is thought to harbor a slim accretion disk 
rather than a standard disk. 
On the other hand, photon indices of the PL-like ULXs 
($\Gamma > 2.0$; e.g., \cite{IC342_VHS}) are steeper than 
those of the classical low/hard state BHBs 
($\Gamma = 1.5$ -- $2.0$; e.g, \cite{Index_in_low/hard}).
Because such a steep photon index is observed from Galactic BHBs
when they are in the very high state (VHS; \cite{GX339_VHS}) 
in which Comptonization plays an important role \citep{XTEJ1550},
the PL-like ULXs are thought to be in the VHS  \citep{IC342_VHS}. 
In the recent Suzaku \citep{Suzaku} observation 
of a nearby normal galaxy NGC 1313, 
\citet{NGC1313_Suzaku} have reinforced 
the VHS and slim disk state interpretations of two ULXs therein, 
X1 and X2, respectively.  

So far, these typical ULX properties have been observed from objects
with very high luminosities (e.g., $\gtrsim 10^{40}$ ergs s$^{-1}$),
which require either a high BH mass or an extreme super-Eddington condition
(or both).  
If, however, such ULX properties are confirmed in less luminous objects 
with a luminosity within a few times the Eddington limit 
of Galactic stellar-mass BHBs,
we will become sure that the ULX phenomena can appear  
without highly-supper Eddington luminosities. 
Particularly important is to search for spectral transitions
among such objects. 

Source 3 (hereafter, Src 3) in the nearby spiral (SABcd) galaxy NGC 2403
is one of the most suitable targets for our purpose. 
Originally discovered by the Einstein observatory \citep{ULX_Einstein},
it was later revealed with ASCA to exhibit an MCD-type spectrum 
with an innermost disk temperature of $T_{\rm in} = 1.10_{-0.09}^{+0.10}$ keV
and an innermost disk radius of $R_{\rm in} = 130$ km \citep{NGC2403_ASCA}.
Furthermore, its bolometric luminosity,
$L_{\rm bol} \sim 2 \times 10^{39} $ ergs s$^{-1}$,
places it in between luminous ULX and ordinary stellar-mass BHBs. 
In order to examine whether this object exhibits the ULX-like behavior, 
we conducted a Suzaku observation of NGC 2403;
then the object was again found in the MCD-like state. 
We also utilized Chandra and XMM-Newton archival data of Src 3,
and discovered a spectral transition that resembles those observed in 
more luminous ULXs.  
In the present paper, 
these results are utilized to strengthen the classification of Src 3
as a ULX with a rather modest luminosity. 
We employ the distance to NGC 2403 of $3.2\pm0.4$ Mpc \citep{NGC2403_distance},
as determined from observations of Cepheid variables. 

\section{Suzaku Observation and Data Reduction} 
\label{sec:suzaku_obs}
The Suzaku observation of NGC~2403 was conducted on 2006 March 16--17, 
during the science working group phase. 
The X-ray Imaging spectrometer (XIS; \cite{XISpaper})
and the Hard X-ray Detector (HXD; \cite{HXDdesign,HXDperform})
onboard Suzaku were operated 
in the normal clocking mode without any window option, 
and in the normal mode, respectively.
The nucleus of the NGC 2403 galaxy was placed
at the XIS nominal position \citep{XRTpaper}. 

Utilizing the HEADAS 6.5.1 software package, 
we reprocessed the data from the Revision 2 processing,
and created new cleaned event files.  
The present paper concentrates on the XIS data, 
because no significant X-ray signals were detected with the HXD 
in this observation. 
We screened the data under the following criteria;
the spacecraft is outside the south Atlantic anomaly (SAA),
the time after an exit from the SAA is larger than 436 s,
the geometric cut-off rigidity is higher than $6$ GV, 
the source elevation above the rim of bright and night Earth is 
higher than \timeform{20D} and \timeform{5D}, respectively, 
and the XIS data are free from telemetry saturation. 
As a result, we have obtained $62.7$ ks of good exposure.
In the scientific analysis below,
we utilize only those events with a grade of 0, 2, 3, 4, or 6.

\section{Results} 
\label{sec:results}
\subsection{X-ray Image} 
\label{sec:suzaku_image}
Figure \ref{fig:img_Suzaku} shows 
a 0.5 -- 10 keV Suzaku XIS image of NGC 2403,
superposed on an optical image taken from 
the Digitized Sky Survey (DSS; \cite{dss}).
Two bright point-like X-ray sources are clearly seen in the figure;
the brighter and fainter of them correspond to Src 3 and Source 5
(hereafter Src~5; \cite{ULX_Einstein}), respectively. 

In addition to the point sources, 
we clearly observe faint X-ray emission, 
extending along the optical galaxy. 
The detailed properties of this extended (possibly diffuse) X-ray emission 
will be reported in a separate paper \citep{NGC2403_diffuse}. 
In the summer of 2004, a supernova, SN2004dj, 
exploded in this galaxy \citep{sn2004dj}.
Although X-ray emission from SN2004dj was detected with Chandra 
in 2004 \citep{sn2004dj_chandra} 
at a $0.5$ -- $8$ keV flux of $1.2 \times 10^{-13}$ ergs s$^{-1}$ cm$^{-2}$, 
we found no significant X-ray enhancement 
at the location of the supernova 
(the filled star in figure \ref{fig:img_Suzaku})
above a $0.5$ -- $8$ keV flux upper limit of 
$1 \times 10^{-14}$ ergs s$^{-1}$ cm$^{-2}$. 

\subsection{X-ray spectrum of NGC~2403 Src~3}  
We integrated X-ray signals of Src 3 within a circle 
denoted as {\bf Src3} in figure \ref{fig:img_Suzaku}. 
The background spectrum was derived within a circle {\bf BGD3}.
In order to reduce possible contributions from the extended X-ray emission,
we adopted a radius of \timeform{2'} 
(corresponding to $1.86$ kpc at the distance of $3.2$ Mpc)
for these regions,
which is slightly smaller than a typical integration region  
used in the Suzaku analysis of point sources ($\sim$\timeform{3'}). 

Figure \ref{fig:lc_Src3_Suzaku} shows 
background-subtracted two-band X-ray lightcurves of Src 3, 
obtained with the three front-illuminated (FI) CCD cameras 
(XIS 0, 2 and 3; \cite{XISpaper}) of the XIS. 
Summed over the three FI CCDs, 
the time-averaged count rate of Src 3~is measured to be 
$0.112\pm 0.002$ cts s$^{-1}$ and $0.078 \pm 0.002$ cts s$^{-1}$
in the soft ($0.5$ -- $2$ keV) and hard ($2$ -- $10$ keV) bands, respectively. 
The data suggest a variation 
on a time-scale of $\gtrsim 10$ ks in the middle of the observation, 
it is statistically insignificant 
($\chi^2/{\rm d.o.f.} = 10.2 / 14$ and $11.2/14$ 
for the soft and hard band, respectively). 
As a result, 
the hardness is found to be rather constant throughout the exposure. 
Below, we hence analyze the time-averaged spectrum of Src 3.

Figure \ref{fig:Src3_Suzaku} shows background-subtracted 
XIS spectra of NGC 2403 Src 3, 
presented without removing the instrumental response. 
The X-ray spectra obtained with FI and background-illuminated (BI) CCD camera
(XIS 1; \cite{XISpaper}) were both binned into energy intervals 
each with at least 100 events.
Thus, the X-ray signals were significantly detected over the 0.4--10 keV range.
The spectrum appears to be rather featureless,
without any clear sign of emission or absorption features.

To analyze the spectra, 
we calculated a response matrix function (rmf)
and an auxiliary response file (arf), 
using {\tt xisrmfgen} and {\tt xissimarfgen} \citep{xissimarf}, respectively. 
The reduction of the effective area 
due to contaminants on the optical blocking filters
of the XIS was taken into account by {\tt xissimarfgen}. 
We performed the spectral fitting using XSPEC version 12.4.0. 

We fitted the Src 3 spectra with the MCD and PL models, 
as a standard way in analyzing the spectra of Galactic BHBs and ULXs.
The FI and BI spectra were jointly fitted,
with the model normalization allowed to differ between them. 
We found that the fluxes from BI and FI spectra agreed 
with each other, within 4\%.

Conducting the spectral fitting, 
we found the absorption column density, which was left free, 
to fall below the Galactic line-of-sight value of 
$N_{\rm H} = 4.1 \times 10^{20}$ cm$^{-2}$
toward NGC 2403 \citep{NH}.
This effect may be  ascribed to residual soft diffuse emission 
within the Src 3 region, 
which remains even after subtracting signals from the BGD3 region,
due to brightness gradients. 
According to \citet{NGC2403_diffuse},
the diffuse emission in NGC 2403 can be described 
by thermal plasma emission model 
with two temperatures of $\sim0.26$ keV and  $\sim0.74$ keV,
having a metalicity of $\sim0.44$ solar.
We hence took into account the contamination, 
by adding two APEC components in XSPEC
with the temperature fixed at $kT = 0.26$ keV and $0.74$ keV,
both of which were subjected to  the Galactic absorption.
Their relative normalization was fixed at the value determined 
with the total diffuse emission (\cite{NGC2403_diffuse}),
while their summed luminosity was left free. 
We left free the absorption to the ULX component. 

The spectral parameters obtained in this analysis 
are summarised in table \ref{table:Src3_Suzaku}. 
The PL model was rejected with $\chi^2 = 349.4/174$,
because the observed X-ray spectrum has a more convex shape,
as revealed by residuals in figure \ref{fig:Src3_Suzaku}. 
On the other hand, 
the MCD model successfully described the X-ray spectrum 
of Src 3 with $\chi^2 = 182.0/174$.
The derived absorption column density for the ULX component,
$N_{\rm H} = 1.40_{-0.42}^{+0.45} \times 10^{21}$ cm$^{-2}$,
is higher by a factor of about 3 than the Galactic value, 
but this amount of excess absorption is attributed reasonably 
to that within the NGC 2403 galaxy, 
and/or an additional absorber localized around the source. 
The thermal plasma components 
were found to have only a negligible contribution to the observed spectrum,  
with a $0.5$ -- $10$ keV luminosity 
of $L_{\rm th} = 3.5 \times 10^{37}$ ergs s$^{-1}$.  
This luminosity is reasonable in view of the results by 
\citet{NGC2403_diffuse}. 

The innermost disk temperature of Src 3 
was obtained as $T_{\rm in} = 1.08_{-0.03}^{+0.02} $ keV.
The X-ray flux was measured to be 
$1.1 \times 10^{-12}$ ergs cm$^{-2}$ s$^{-1}$ in the 0.7 -- 7 keV range,
without correcting for absorption.
These give an absorption-corrected bolometric luminosity of 
$L_{\rm bol} = 1.82 \times 10^{39} \alpha $ ergs s$^{-1}$,
with $\alpha = (\cos 60^\circ /\cos i)$ 
where $i$ is the inclination of the accretion disk to our line of sight.
The value of $T_{\rm in}$ agrees with that obtained 
with ASCA in 1997 \citep{NGC2403_ASCA} within statistical errors, 
while  $L_{\rm bol}$ is lower by about $20$\%.  
Through a relation $L_{\rm bol} = 4\pi \sigma r_{\rm in}^2 T_{\rm in}^4$, 
with $\sigma$ being the Stefan-Boltzmann constant,
the apparent innermost disk radius is calculated 
as $r_{\rm in} =102.6_{-5.8}^{+6.4}$ km. 
The true innermost radius is determined as
$R_{\rm in} = 122.1_{-6.8}^{+7.7}~\alpha^{1/2}$ km, 
applying the correction as $R_{\rm in } = \xi \kappa^2 r_{\rm in}$,
where $\xi = 0.412$ is a correction factor 
for the inner boundary condition \citep{xi}, 
and 
$\kappa = 1.7$ is a spectral hardening factor \citep{kappa}
which represents the ratio of the color temperature
to the effective temperature. 

We also examined the ``variable-$p$'' disk model,
in which the dependence of the local temperature on the radius $r$ 
from the BH is assumed to scale as $r^{-p}$, 
with the index $p$ being a positive free parameter
\citep{pfree_disk}. 
While this model with $p = 0.75$ reduces to a simple MCD model,
that with smaller value of $p$, down to 0.5,
is considered to approximate the X-ray spectra from a slim disk 
\citep{slimdisk,XTEJ1550,ULX_asca2,NGC1313_Suzaku}.
The fit was acceptable,
but with no significant improvement ($\chi^2 = 181.9/173$) 
over the MCD fit.
All the parameters became consistent with those of the MCD model;
$p = 0.73_{-0.07}^{+0.12}$, $T_{\rm in} = 1.09 \pm 0.07 $ keV,  
and $R_{\rm in} = 116.4_{-28.7}^{+39.4}$ km.

\section{Analysis of Chandra and XMM-Newton Data} 
In order to better understand the nature of Src 3,
its intensity-correlated spectral variations 
are considered to be of particular importance (e.g, \cite{ULX_asca2}). 
Therefore, we analyzed the archival Chandra~and XMM-Newton~data of NGC 2403. 

\subsection{Chandra observations} 
\label{sec:chandra_osb}
So far, 5 Chandra ACIS exposures toward NGC 2403 have been conducted,
4 of which were motivated by the discovery of SN2004dj \citep{sn2004dj}. 
However, NGC 2403 Src 3 was within the ACIS field of view 
on only three occasions. 
Furthermore, 
in the first observation performed on 2001 April 17 (ObsID = 2014), 
the ACIS spectrum of NGC 2403 Src 3 was severely distorted 
by event pileup ($\gtrsim 30 \%$; \cite{NGC2403_Chandra}).
Therefore, we analyzed only the data 
in the remaining two observations (ObsID = 4628 and 4630),
wherein Src 3 was placed at a relatively large distance from the ACIS aimpoint
(see $\Delta \theta$ in table \ref{table:chandra}),
so that a broadened point spread function made the event pileup insignificant. 
The log of these two observations is shown in table \ref{table:chandra}.
The data were read out in the timed-exposure mode 
with the standard frame time of $3.2$ s, and 
telemetered in the FAINT format. 
In ObsID = 4628 and 4630, 
Src 3 was placed on ACIS-S3 (a BI chip) and S2 (an FI one),
respectively.  

Utilizing the CIAO 3.4 software and referring to CALDB 3.3.0.1,
we reprocessed all the data to create new level-2 event files
in the standard manner.
We removed {\tt acis\_detect\_afterglow} correction, and 
applied a ``new'' bad pixel file created by {\tt acis\_run\_hotpix} 
\footnote{http://asc.harvard.edu/ciao3.4/guides/acis\_data.html}. 
After removing X-ray point sources detected by {\tt wavdetect} 
within the ACIS field of view, 
we produced a 0.3 - 12 keV lightcurve covering the entire CCD chip,
and then discarded those time intervals where  
the count rate exceeds 120\% of that averaged during the observation.
This procedure gave us the good exposures 
as listed in table \ref{table:chandra}. 
The ACIS spectra of NGC 2403 Src 3 were extracted  
within a circle centered on the X-ray peak, 
and those of BGD from a concentric annulus.
The radii of these regions are also shown 
in table \ref{table:chandra}. 
The rmf and arf files were generated, assuming an X-ray point source,  
with the CIAO tools {\tt mkacisrmf} and {\tt mkarf}, respectively.

\subsection{XMM-Newton observations} 
\label{sec:newton:obs}
So far, XMM-Newton observations of NGC 2403 have been performed 5 times.
However, 
only the data obtained in three of them are public at present,
of which the observational log is given in table \ref{table:xmm}. 
Although the results in the individual observations 
were separately reported by several authors 
\citep{ULX_XMM,ULX_XMM1,ULX_XMM2}, 
we re-analyzed all the data in a sytematic way. 
The third observation (ObsID = 0164560901) was motivated 
by the explosion of SN2004dj.
On all these occasions,
the EPIC MOS and pn cameras were operated in the nominal full frame mode. 
The medium optical-blocking filter was adopted 
in the first and second observations 
(ObsID = 0150651101 and 0150651201, respectively),
while the thick one was utilized in the third observation. 
In the first and second observations, we do not use the pn data,
because Src 3 fell on the gap of the pn chips.

We used the Science Analysis System (SAS) version 7.0.0 
to reduce the XMM-Newton data.
All data were reprocessed by {\tt emchain} or {\tt epchain}, 
based on the Current Calibration Files (CCF), 
which was latest at the beginning of our analysis (2007 July). 
After \citet{XMM_BGD}, 
we rejected high-background periods,
using the point-source removed lightcurve 
with PATTERN == 0 and \#XMMEA\_EM/P in 10 -- 15 keV. 
As a result, we obtained good exposures, 
as shown in table \ref{table:xmm}. 
For the spectral analysis,
MOS events with PATTERN $\le 12$, \#XMMEA\_EM and FLAG = 0, 
and pn ones with PATTERN $\le 4$, \#XMMEA\_EP and FLAG = 0,
were finally selected.  

We integrated the EPIC spectra of NGC 2403 Src 3 and BGD, 
within a circle of $36''$ radius and 
an annulus of $72''$--$108''$ radius, respectively, 
both centered on Src 3. 
Because the pn image of the third observation revealed 
a faint point-like source within the BGD annulus,
a circle of $30''$ radius around this contaminating source 
was removed from both MOS and pn data.   
We created the rmf and arf,
utilizing the SAS tools, {\tt rmfgen} and {\tt arfgen} respectively. 

\subsection{Spectral modeling}       
Figure \ref{fig:ufspec} compares the Chandra and XMM-Newton spectra 
of NGC 2403 Src 3 with the Suzaku one, 
in the form of the unfolded spectrum (panel a)
and of the ratio to the Suzaku best-fit MCD model (panel b).
All the data are presented in 0.7 -- 7 keV, 
where the energy ranges of all the three instruments overlap. 
In the Chandra (4628) and XMM-Newton (0164560901) observations, 
the spectral shape of Src 3 thus coincides 
well with those in the Suzaku~observation,
indicating that the source was 
in the MCD-like spectral state on these occasions. 
In addition, the spectral normalization is similar, 
to within $\sim 20\%$, among these three data sets. 
In contrast, 
the X-ray spectrum in the Chandra observation of ObsID = 4630 
exhibits a distinct and less convex shape, 
which is approximated by a power-law, at least above $\sim 1$ keV
where the effect of interstellar absorption become unimportant.
The flux on this occasion is estimated to be lower by about $15\%$ 
than that in the Suzaku~observation.   
In figure \ref{fig:ufspec}, 
the deconvolved spectra have been derived using the absorbed MCD or PL model
(whichever is more successfull) to be described below.

The raw Chandra and XMM-Newton spectra of Src 3 
in the individual observations are
shown in figures \ref{fig:Src3_Chandra} and \ref{fig:Src3_Newton}, 
respectively.
We applied the MCD and PL models to these spectra,
in the same way as for the Suzaku~data,  
and obtained the results summarised in table \ref{table:Src3}.
We neglected the contamination from the diffuse X-ray emission 
associated with the host galaxy, 
because the integration area,
we adopted for Chandra and XMM-Newton data,
were more than an order of magnitude smaller than 
that for Suzaku data. 
As figure \ref{fig:ufspec} suggests,
the MCD model better reproduced the observed X-ray spectra in general,
except for that from the Chandra observation of ObsID = 4630. 
When the source was in the MCD-like state,
the spectral parameters exhibited only small deviations 
from the Suzaku measurements. 
The bolometric luminosity of Src 3
was found to vary only about 10\% among these observations. 
Similar MCD parameters were reported by \citet{ULX_XMM} and \citet{ULX_XMM1},
based on the third XMM-Newton observation. 

The Chandra spectrum of ObsID = 4630, in contrast, 
was better reproduced by a PL model 
with a photon index of $\Gamma = 2.37 \pm 0.08$. 
The X-ray flux on this occasion,
$9.3 \times 10^{-13}$ ergs cm$^{-2}$ s$^{-1}$  in the 0.7 -- 7 keV range, 
was lowest among the data sets analyzed here. 
The absorption column density became larger by a factor of about $2$ -- $3$
than those in the other observations;
$N_{\rm H} = 3.73_{-0.33}^{+0.35} \times 10^{21}$ cm$^{-2}$.  
Similarly, an apparent $N_{\rm H}$ change was reported 
from a transient ULX, CXOU J132518.2-430304, 
which was discovered 
within the host galaxy of Centaurus A \citep{ULX_in_CenA}. 
However, such significant changes in $N_{\rm H}$, 
associated with transitions between the PL-like and MCD-like states, 
are uncommon among Galactic BHBs. 
Therefore, this $N_{\rm H}$ change from Src 3 is likely to be an artifact, 
caused presumably because a simple PL model is inappropriate to describe 
the softest end of the Chandra spectrum from ObsID=4630.
We discuss this issue briefly in section 5.

Although the MCD model gave a better fit than the PL model 
to all the XMM-Newton data, 
the third XMM-Newton dataset still exhibit some residuals 
with $\chi^2/{\rm d.o.f.} = 637.9/568$ (figure 6c).
This urged us to apply the variable-$p$ disk model 
to this dataset and the other MCD-like ones, 
as we did for the Suzaku spectrum.  
Then, all the spectra in the MCD-like states 
were successfully reproduced by the variable-$p$ disk model.
The fit was significantly improved in the third XMM-Newton observation 
and the Chandra ObsID $=$ 4628, and yielded 
$p = 0.58_{-0.02}^{+0.03}$ and $0.60_{-0.05}^{+0.07}$, respectively.  

\section{Discussion}  
\label{sec:discussion}
In the Suzaku observation of NGC 2403 performed on 2006 March 16 -- 17, 
Src 3 exhibited a curved 0.4 -- 10 keV spectrum,
which can be reproduced by the MCD model.
We obtained the innermost disk temperature and radius of the source 
as $T_{\rm in} = 1.08_{-0.03}^{+0.02} $ keV
and $R_{\rm in} = 122.1_{-6.8}^{+7.7}~\alpha^{1/2}$ km, respectively. 
The bolometric luminosity of the source was measured as 
$L_{\rm bol} = 1.82 \times 10^{39} \alpha$ ergs s$^{-1}$. 
Making use of the archival Chandra and XMM-Newton data, 
we found the source to exhibit similar MCD-type spectrum on 4 occasions, 
while a PL-shaped spectrum on one occasion (Chandra observation of ObsID = 4630)
when the 0.7 -- 7 keV flux was about $15$\% and $17$\% lower than 
those in the Suzaku observation 
and
the Chandra observation of ObsID = 4628, respectively. 
This flux change between the MCD-like and PL-like states are 
larger than the relative flux uncertainties between 
Suzaku, Chandra, and XMM-Newton 
($\lesssim 10$\%; e.g., \cite{MCG-5-23-16,cross_cal2,cross_cal}).

In the MCD-like states, $T_{\rm in }$ and $L_{\rm bol}$ of Src 3 
exhibited only small variations across more than 8 years 
from the ASCA observation \citep{NGC2403_ASCA} to the Suzaku one.
Both of these parameters are located 
at the lowest end of those of typical ULXs.
The value of $R_{\rm in}$ derived in these observations corresponds to 
the radius of the last stable orbit,
$3 R_{\rm S}$ ($R_{\rm s}$ being the Schwarzschild radius), 
around a non-rotating BH with a mass of $\sim 13 \alpha^{1/2} M_\odot $. 
Because the Eddington luminosity of such a BH becomes 
$L_{\rm E} \sim 2 \times 10^{39} \alpha^{1/2} $ ergs s$^{-1}$,    
the properties of Src 3 in the MCD-like states can be understood consitently
 as a $\sim 13 M_\odot$ BH radiating at close to $L_{\rm E}$,  
wherein a standard accretion disk is formed down to $\sim 3 R_{\rm S}$,
like in the high/soft state of normal Galactic BHBs.
This apparently reconfirms the conclusion by \citet{NGC2403_ASCA}. 
Then, can we really explain the overall behavior of Src 3, 
by assuming that it was in the standard high/soft state 
during the Suzaku observation ?

One difficulty with the high/soft state interpretation 
comes from the fact that the third XMM-Newton dataset 
preferred a variable-$p$ disk model with $p = 0.58_{-0.02}^{+0.03}$,
which is smaller than that in the standard accretion disk.
To see this in a systematic manner, we plotted in figure \ref{fig:variable-p}, 
the parameters determined from the variable-$p$ disk model,
against the observed X-ray flux in the $0.7$ -- $7$ keV range. 
Although errors are rather large, 
the plot reveals a tendency that $R_{\rm in}$ and $p$ become smaller,
and $T_{\rm in}$ gets higher, as the flux increases.
Such behavior is numerically predicted by the slim disk model 
(e.g., \cite{slimdisk,limit_cycle}).
The correlation is not affected significantly by possible flux uncertainties 
among the three sattelites, 
because the three XMM-Newton data points reveal the same behavior. 
Therefore, it is natural to regard 
that the source was actually in the slim disk state during these observations.
Moerover, figure \ref{fig:variable-p} indicates that 
Src 3 was in the transition regime 
between the standard high/soft state ($p = 0.75$)
and 
the slim disk regime ($p \le 0.75$) when observed with Suzaku.

Another difficulty with the high/soft interpretation is provided 
by the spectral transition, revealed by the Chandra data (ObsID = 4630),
when the source was  $\sim 15$\% less luminous than in the Suzaku data. 
The spectral slope of this PL-like state, $\Gamma = 2.37 \pm 0.08 $,
is too steep for the  source to be interpreted 
as in the low/hard state 
($\Gamma = 1.5$ -- $2.0$; e.g, \cite{Index_in_low/hard}), 
which would be realized 
when the source becomes less luminous than in the high/soft state. 
In contrast, this steep slope agrees very well with 
those of Galactic BHBs in the VHS \citep{GX339_VHS,XTEJ1550_VHS}.
Furthermore, Galactic BHBs have been observed to become less luminous 
when they make a transition from the slim disk state to the VHS 
\citep{XTEJ1550,4U1630‑47}. 
These comparisons suggest 
that the PL-like state of Src 3 corresponds to the VHS. 

Since Comptonization of disk photons by hot coronae is thought to become
important in the VHS \citep{XTEJ1550}, 
we tried a Comptonization model, {\bf compTT} in XSPEC \citep{comptt}, 
on the Chandra spectrum of Src 3 in ObsID = 4630.
The electron temperature of the Comptonizing coronae 
was fixed at $T_{\rm e} = 20 $ keV, 
as seen in the VHS of XTE J1550-564 \citep{XTEJ1550_VHS}.
The model successfully described the data  
with parameters shown in table \ref{table:compton_4630},
giving a fit goodness similar to that with the PL model.
The optical depth of the corona, $\tau = 1.2 \pm 0.1$,
is comparable to those of the Galactic BHB in the VHS 
($\tau \sim 2$; e.g,  \cite{GROJ1655_VHS,XTEJ1550_VHS}).
Furthermore, the value of $N_{\rm H}$ has become consistent 
with those found in the MCD-like states; 
this is because the Comptonized continuum flattens 
in energies below that of the seed photons ($T_{\rm seed} \sim 0.25$ keV). 
Although the derived value of $T_{\rm seed}$ is lower than the disk temperature 
$T_{\rm in} \sim  1.1$ keV measured in the MCD-like state, 
the discrepancy could be due to a disk truncation at a radius
larger than that of the last stable orbit,  
as is suggested in the ``strong'' VHS of XTE J1550-564 \citep{XTEJ1550_VHS}.

From these considerations,
we conclude that NGC 2403 Src 3 normally resides 
at the lowest end of the slim disk state
and makes occasional transitions into the VHS.  
Since the dominance of these two spectral states 
is regarded as an essential property of the ULXs 
(e.g., \cite{IC342_VHS,M81X9,NGC1313_Suzaku}),
we may regarded Src 3 as a ULX, in spite of its moderate luminosity. 
In addition, we may regard Src 3 as radiating at $\sim L_{\rm E}$,
for the following two reasons. 
One is that the transisions of BHBs 
between the VHS and the slim disk state is 
obvseved at $\gtrsim 0.3 L_{\rm E}$ \citep{XTEJ1550,4U1630‑47},
while the other is that the value of $p \sim 0.75$ would not be observed 
if the source were radiating with $\gg L_{\rm E}$,
Thus, $L_{\rm bol}$ of the source gives the BH mass 
as $M \sim 10$ -- $20 M_\odot$. 

We may independently estimate the BH mass 
using $R_{\rm in}$ determined by the variable-$p$ disk model.
In this case, the simple assumption of $R_{\rm in} = 3 R_{\rm S}$ 
would underestimate a mass of the BH that hosts a slim disk,
because the X-ray photons would also be radiated
from the region inside the last stable orbit \citep{slimdisk}.
Recently, 
\citet{mass_in_slim} evaluated a factor to correct the underestimation
as $1.2$ -- $1.6$,
by analyzing numerically simulated X-ray spectra
from the slim disk with the variable-$p$ disk model
(the extended disk black body model in their paper) 
for a wide range of mass accretion rate.
Combining this correction factor and 
$R_{\rm in}$ derived from our variable-$p$ disk model fit 
(figure \ref{fig:variable-p}b),
we estimate the BH mass of NGC 2403 Src 3 as $M = 9$ -- $15 M_\odot$.
This value agrees very well with that derived 
from the Eddington limit argument. 
Therefore,  we can consistently interpret the available data of NGC 2403 Src 3,
by considerig that it hosts a BH with a mass of $M = 10$ -- $15 M_\odot$,
radiating at $\sim L_{\rm E}$.   

Finally, let us discuss implications of the present results 
on the ULX phenomenon in general. 
Obviously, they do not give direct support 
to the intermediate mass BH interpretation of ULX, 
since the observed luminosity of Src 3 is well 
within the Eddington limits of ordinary-mass stellar BH. 
Nevertheless, one important discovery is 
that this relatively low-luminosity object exhibits clear ULX behavior, 
including in particular a transition between the slim disk state and the VHS.
In other words, the ULX phenomenon appears 
even at a luminosity range of $\sim 2 \times 10^{39}$ ergs s$^{-1}$. 
This is within $\sim 3$ times the Eddington luminosity of any stellar-mass BHs,
even if we consider the least massive known BHBs 
as GRO J0422$+$32 ($3.97 \pm 0.95 M_\odot$; \cite{GROJ0422_mass}),
GRO J1655$-$40 ($5.5$--$7.9 M_\odot$; \cite{GROJ1655_mass}),
and GX 339-4 ($5.8 \pm 0.5 M_\odot$; \cite{GX339_mass}).
Consequently, the ULX behaviour in this case does not require 
any highly  ($\sim 10$) super-Eddington luminosity. 
Extrapolating this conclusion to more luminous ULXs, 
we suggest that they are not radiating 
at highly super-Eddington luminosities,
either, even though their mass accretion rates are 
very likely to be "super-critical".
That is, ULXs are thought to be BHs 
which are significantly more massive than Galactic BHBs,
accreting matter at super-critical rates, 
and yet radiating at about their Eddington limit. 
\\

We are grateful to all the members of the Suzaku team,
for the successful operation and calibration.
We also thank Dr. Ohsuga for his valuable theoretical comments.
This research has made use of the archival Chandra~data and
its related software provided by the Chandra~X-ray Center (CXC).
A part of the result is based on the observation obtained with XMM-Newton,
an ESA science mission with instruments and contributions
directly funded by ESA Member States and NASA.
We have made extensive use of the NASA/IPAC Extra galactic Database
(NED; the Jet Propulsion Laboratory, California Institute
of Technology, the National Aeronautics and Space Administration).
The optical DSS image of NGC 2403~was downloaded from SkyView
\footnote{http://skyview.gsfc.nasa.gov/}.
The ROSAT source catalog were taken 
from the ROSAT Source Browser by Dr. Englhauser
\footnote{http://www.xray.mpe.mpg.de/cgi-bin/rosat/src-browser}.



\clearpage 
\begin{table}[htbp]
\caption{Best-fit parameters to the Suzaku spectra of Src 3.}
\label{table:Src3_Suzaku}
\begin{center}
\begin{tabular}{llll}
\hline\hline 
Model                                   &  MCD                  & variable-$p$          & PL                  \\
\hline       
$N_{\rm H}$ ($10^{21}$ cm$^{-2}$)         &  $1.40_{-0.42}^{+0.45}$ & $1.52_{-0.74}^{+1.02}$  & $8.66_{-0.65}^{+0.69}$ \\
$T_{\rm in}$ (keV)                       &  $1.08_{-0.03}^{+0.02}$ & $1.09 \pm 0.07 $      & --                   \\
$p$, $\Gamma$ \footnotemark[$*$]        &  --                   & $0.73_{-0.07}^{+0.12}$  & $2.79 \pm 0.07$      \\
$R_{\rm in}$ (km)                        &  $122.1_{-6.8}^{+7.7}$  & $116.4_{-28.7}^{+39.4}$ & --                   \\
$L$ \footnotemark[$\dagger$]            &  $1.82$               & $1.89$                & $4.10$               \\
$L_{\rm th}$  \footnotemark[$\ddagger$]  &  $0.035$              & $0.038$               & $0.174$               \\
\hline       
$\chi^2/{\rm d.o.f.}$                    &  $182.0/174$          & $181.9/173$           & $349.4/174$         \\
\hline       
\multicolumn{4}{@{}l@{}}{\hbox to 0pt{\parbox{90mm}{\footnotesize
\footnotemark[$*$] Index $p$ of radial temperature dependence for the variable-p model, or
                   the photon index $\Gamma$ for the PL one.
\par\noindent
\footnotemark[$\dagger$] $L_{\rm bol}$  for the MCD and variable-p model, 
                         or absorption-corrected $0.5$ -- $10$ keV luminosity for the PL model, 
                         both in $10^{39}$ ergs s$^{-1}$.
\par\noindent
\footnotemark[$\ddagger$] Absorption-corrected $0.5$ -- $10$ keV luminosity 
                          of the sum of the two APEC components, in $10^{39}$ ergs s$^{-1}$.
}\hss}}
\end{tabular}
\end{center}
\end{table}

\begin{table}[h]
\caption{Log of archival Chandra observations of NGC 2403, utilized in the present paper.}
\label{table:chandra} 
\begin{center}
\begin{tabular}{lll}
\hline\hline 
ObsID           & 4628                  & 4630 \\ 
\hline 
Date            & 2004 Aug 23           & 2004 Dec 22   \\
Exposure (ks)   & 46.5                  &48.7           \\
Mode            & \multicolumn{2}{c}{Timed Exposure} \\
Frame Time      & \multicolumn{2}{c}{$3.2$ s} \\       
Format          & \multicolumn{2}{c}{FAINT} \\
ACIS Chip\footnotemark[$*$]      
                & S3                    & S2            \\
$\Delta\theta$\footnotemark[$\dagger$]      
                & $5'.8$                & $4'.9$              \\
$r_{\rm Src 3}$\footnotemark[$\ddagger$]        
                & $15''.7$              & $13''.8$              \\
$r_{\rm BGD}$\footnotemark[$\S$]  
                & $19''.6$--$23''.6$    & $17''.7$--$21''.6$               \\
Signal (cts s$^{-1}$) \footnotemark[$\|$]     
                & 0.16                  & 0.11                 \\
\hline 
\multicolumn{3}{@{}l@{}}{\hbox to 0pt{\parbox{80mm}{\footnotesize
\par\noindent
\footnotemark[$*$] ACIS CCD chip on which Src 3 is located.
\par\noindent
\footnotemark[$\dagger$] Off axis angle of Src 3 from the aim point
\par\noindent
\footnotemark[$\ddagger$] Radius of the source integration circle. 
\par\noindent
\footnotemark[$\S$] Radius of the BGD integration annulus.
\par\noindent
\footnotemark[$\|$] Count rate of Src 3 in 0.3 -- 10 keV 
}\hss}}
\end{tabular}
\end{center}
\end{table}

\begin{longtable}{lllll}
\caption{Log of archival XMM-Newton observations of NGC 2403.}
\label{table:xmm} 
\hline\hline 
ObsID           && 0150651101    & 0150651201    & 0164560901    \\   
\hline 
\endfirsthead   
\hline 
\endhead
\hline 
\endfoot
\multicolumn{5}{@{}l@{}}{\hbox to 0pt{\parbox{90mm}{\footnotesize
\par\noindent
}\hss}}
\endlastfoot
\multicolumn{2}{l}{Date}   & 2003 Apr 30   & 2003 Sep 11   & 2004 Sep 12--13 \\ 
\multicolumn{2}{l}{Mode}   & \multicolumn{3}{c}{Full Frame} \\
\multicolumn{2}{l}{Filter} & Thin          & Thin          & Medium               \\
Exposure (ks)                   &MOS1           & 5.26          & 6.98          & 57.1 \\
                           &MOS2           & 5.52          & 7.20          & 56.2 \\
                           &pn             & --            & --            & 49.6 \\
\hline 
\end{longtable}

\begin{longtable}{lllllllll}
\caption{Summary of fitting to the Chandra and XMM-Newton spectra of Src 3}
\label{table:Src3}
\hline\hline 
Instrument & ObsID      & Model      & $N_{\rm H}$             & $T_{\rm in}$                  & $p$, $\Gamma$ \footnotemark[$*$]
                                     & $R_{\rm in}$            & $L$ \footnotemark[$\dagger$] & $\chi^2/{\rm d.o.f.}$      \\
           &            &            & ($10^{21}$ cm$^{-2}$ )  & (keV)                        &
                                     & (km)                   &                              &                 \\          
\hline 
\endfirsthead   
\hline 
\endhead
\hline 
\endfoot
\multicolumn{9}{@{}l@{}}{\hbox to 0pt{\parbox{170mm}{\footnotesize
\par\noindent
\footnotemark[$*$] Index $p$ of radial temperature dependence for the variable-p model, 
                   or the photon index $\Gamma$ for the PL model.
\par\noindent
\footnotemark[$\dagger$] $L_{\rm bol}$ for the MCD and variable-$p$ model, 
                         or absorption-corrected $0.5$ -- $10$ keV luminosity for the PL model, 
                         both in $10^{39}$ ergs s$^{-1}$.
}\hss}}
\endlastfoot
Chandra   & 4628       & MCD        & $1.47_{-0.14}^{+0.15}$   & $1.06 \pm 0.04 $       & --                
                                    & $127.4_{-8.3}^{+8.8}$    & $1.86$                 & $166.9/156$               \\
          &            & variable-$p$ & $2.23 \pm 0.44$        & $1.23_{-0.11}^{+0.15}$   & $0.60_{-0.05}^{+0.07}$ 
                                    & $70.3_{-20.8}^{+28.1}$   & $2.48$                 & $158.4 / 155$             \\
          &            & PL         & $3.83_{-0.24}^{+0.26}$   & --                     & $2.28_{-0.06}^{+0.07} $     
                                    & --                     & $2.62$                 & $230.0/156$               \\
          & 4630       & MCD        & $0.97_{-0.19}^{+0.21}$   & $1.04_{-0.04}^{+0.05}$   & --  
                                    & $115.3_{-9.9}^{+11.0}$   & $1.42$                 & $194.8/123$               \\
          &            & variable-$p$ & $3.55_{-0.32}^{+0.34}$   & $3.3 (> 2.3)$          & $0.46 \pm 0.01 $ 
                                    & $4.19_{-3.87}^{+6.64}$   & $10.4$                 & $109.9/122$               \\
          &            & PL         & $3.73_{-0.33}^{+0.35}$   & --                     & $2.37 \pm 0.08$  
                                    & --                     & $2.14$                 & $110.7/123$               \\   
\hline     
XMM-Newton       & 0150651101 & MCD        & $1.40_{-0.38}^{+0.43}$  & $1.09 \pm 0.10$                        & --
                                     & $116.5_{-19.2}^{+23.0}$ & $1.72$                                 & $31.1/37$                 \\
           &            & variable-$p$ & $1.44_{-0.96}^{+1.31}$  & $1.10_{-0.18}^{+0.34}$                 & $0.74_{-0.18}^{+1.81}$ 
                                     & $112.4_{-68.8}^{+81.9}$ & $1.75$                                 & $31.1/36$                 \\ 
           &            & PL         & $3.77_{-0.66}^{+0.76}$  & --                                     & $2.23_{-0.17}^{+0.18}$
                                     & --                      & $2.42$                                 & $46.3/37$                 \\ 
           & 0150651201 & MCD        & $1.66_{-0.36}^{+0.41}$  & $1.11 \pm 0.09$                        & -- 
                                     & $116.5_{-17.6}^{+20.8}$ & $1.87$                                 & $33.7/45$                 \\
           &            & variable-$p$ & $2.14_{-1.22}^{+1.15}$  & $1.21_{-0.23}^{+0.38}$                 & $0.65_{-0.11}^{+0.40}$
                                     & $83.1_{-48.5}^{+110.4}$ & $2.16$                                 & $33.2/44$                 \\
           &            & PL         & $4.15_{-0.64}^{+0.71}$  & --                                     & $2.22_{-0.15}^{+0.16}$
                                     & --                      & $2.65$                                 & $47.5/45$                 \\ 
           & 0164560901 & MCD        & $1.66 \pm 0.07$         & $1.06 \pm 0.02$                        & -- 
                                     & $130.9_{-4.6}^{+4.8}$   & $1.95$                                 & $637.9/568$               \\
           &            & variable-$p$ & $2.59_{-0.21}^{+0.22}$  & $1.27 \pm 0.06$                        & $0.58_{-0.02}^{+0.03}$
                                     & $64.3_{-10.1}^{+11.1}$  & $2.85$                                 & $585.6/567$               \\
           &            & PL         & $4.30_{-0.12}^{+0.13}$  & --                                     & $2.43 \pm 0.03$ 
                                     & --                      & $2.81$                                 & $943.3/568$               \\ 
\hline     
\end{longtable}

\begin{table}[h]
\caption{Parameters of compTT fitting to the Chandra spectrum of ObsID = 4630.}
\label{table:compton_4630} 
\begin{center}
\begin{tabular}{ll}
\hline\hline 
 Parameters                              & Value            \\
\hline 
$N_{\rm H}$ ($10^{21}$ cm$^{-2}$)          & $1.4_{-0.7}^{+2.6}$ \\ 
$T_{\rm seed}$\footnotemark[$*$] (keV)     & $0.25 (< 0.29)$  \\
$T_{\rm e}$\footnotemark[$\dagger$](keV)  & $20$ (fix)        \\
$\tau$\footnotemark[$\ddagger$]          & $ 1.15 \pm 0.11$  \\
\hline 
$\chi^2/{\rm d.o.f.}$                     & $111.14/122$ \\
\hline 
\multicolumn{2}{@{}l@{}}{\hbox to 0pt{\parbox{80mm}{\footnotesize
\par\noindent
\footnotemark[$*$] Energy of seed soft photons 
\par\noindent
\footnotemark[$\dagger$] Electron temperature of the Comptonizing corona
\par\noindent
\footnotemark[$\ddagger$] Optical depth of the Comptonizing corona
}\hss}}
\end{tabular}
\end{center}
\end{table}

\clearpage 
\begin{figure*}[h]
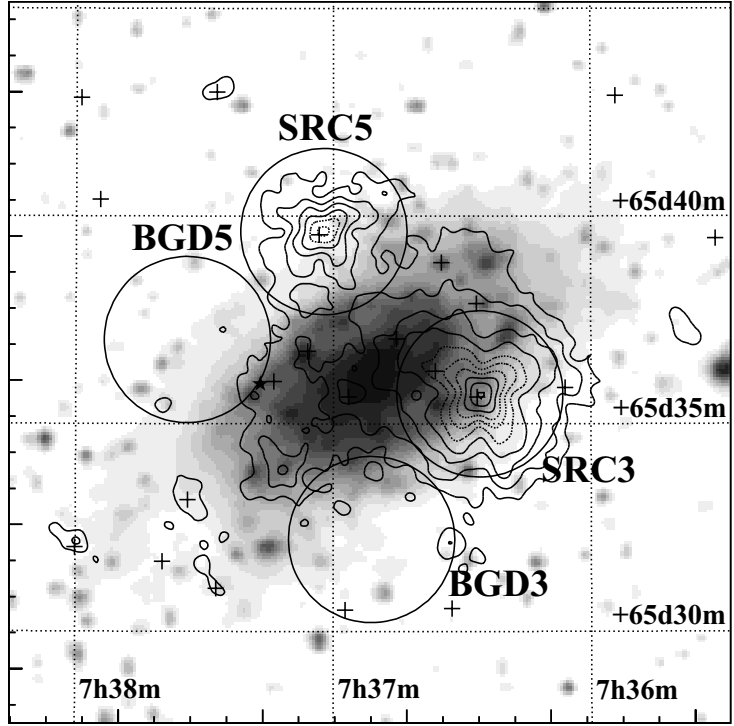

\begin{center}
\FigureFile(120mm,120mm){figure1.ps}
\end{center}
\caption{Suzaku XIS contour image of NGC 2403 in the 0.5 -- 10 keV range,
overlaid on the optical DSS gray-scale image.
Data from all the XIS CCD chip are summed up. 
Neither background-subtraction nor exposure-correction is performed to the image.
The source signals of Src 3 and the associated background are integrated within {\bf Src3} and 
{\bf BGD3},
while those of Src 5 are extracted from {\bf Src5} and {\bf BGD5}.
The positions of the ROSAT X-ray sources, 
taken from the second ROSAT source catalog of pointed observations
with the position sensitive proportional counter (2RXP),  
are indicated with crosses, and that of SN2004dj with the filled star. } 
\label{fig:img_Suzaku}
\end{figure*}

\begin{figure}[h]
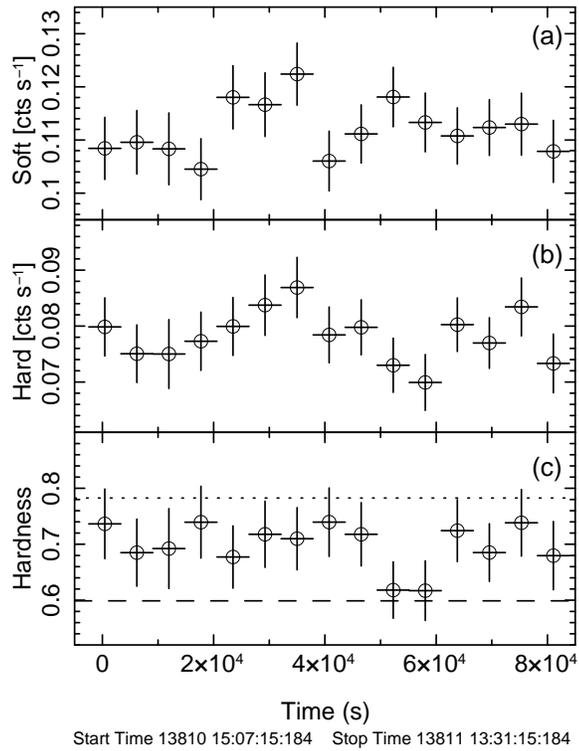

\FigureFile(80mm,80mm){figure2.ps}
\caption{Background-subtracted XIS FI lightcurve of NGC 2403 Src 3.
Each time bin is set to 5760 s, 
corresponding to the orbital period of Suzaku.   
Panels (a) and (b) show the soft (0.5 -- 2 keV) and hard (2 -- 10 keV)
energy band lightcurves, respectively. 
The hardness, simply calculated as the ratio of the hard band count rate 
to the soft band one, is shown in panel (c). 
Dashed and dotted lines in panel (c) indicate predictions by MCD models 
with of $T_{\rm in} = 1.0$ keV and $1.2$ keV, respectively.}
\label{fig:lc_Src3_Suzaku}
\end{figure}

\begin{figure}[h]
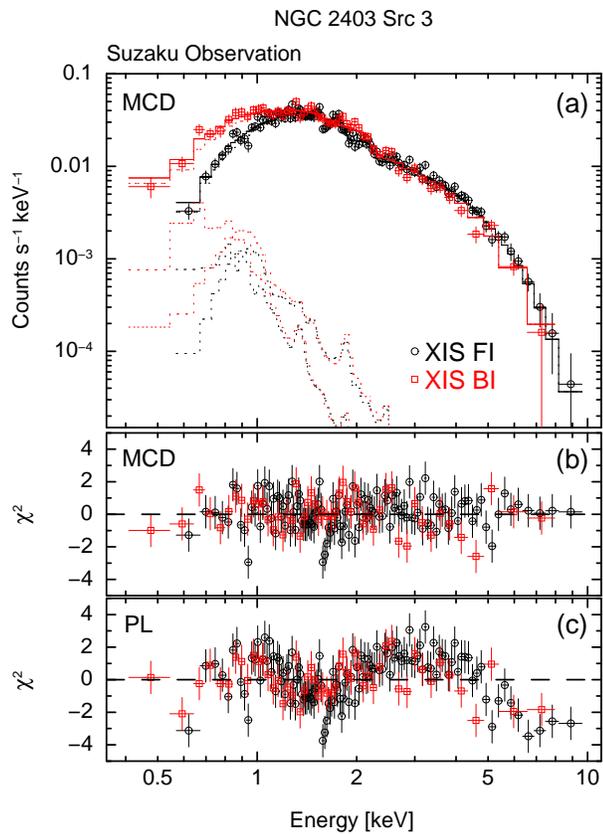

\FigureFile(80mm,80mm){figure3.ps}
\caption{(a) Suzaku XIS spectra of NGC 2403 Src 3, 
shown together with the best-fit MCD model.  
The FI and BI data are shown with black and red points, respectively. 
Two APEC components are also indicated (see text). 
The data are binned into pixels with at least 100 events.
Panels (b) and (c) show the residuals from the MCD and PL models, respectively.}
\label{fig:Src3_Suzaku}
\end{figure}

\begin{figure*}[h]
\FigureFile(80mm,80mm){figure4a.ps}
\FigureFile(80mm,80mm){figure4b.ps}
\caption{(a) Unfolded spectra of NGC 2403 Src 3.
The black, red, green and blue points indicate the Suzaku data, 
the Chandra data of ObsID = 4628, those of ObsID = 4630, 
and the XMM-Newton data of ObsID = 0164560901, respectively. 
Panels (b) - (d) show the Chandra and XMM-Newton spectra, 
divided by the prediction of the best-fit MCD model 
determined by the Suzaku spectrum. 
The color specification is the same as in panel (a)}
\label{fig:ufspec}
\end{figure*}

\begin{figure*}[h]
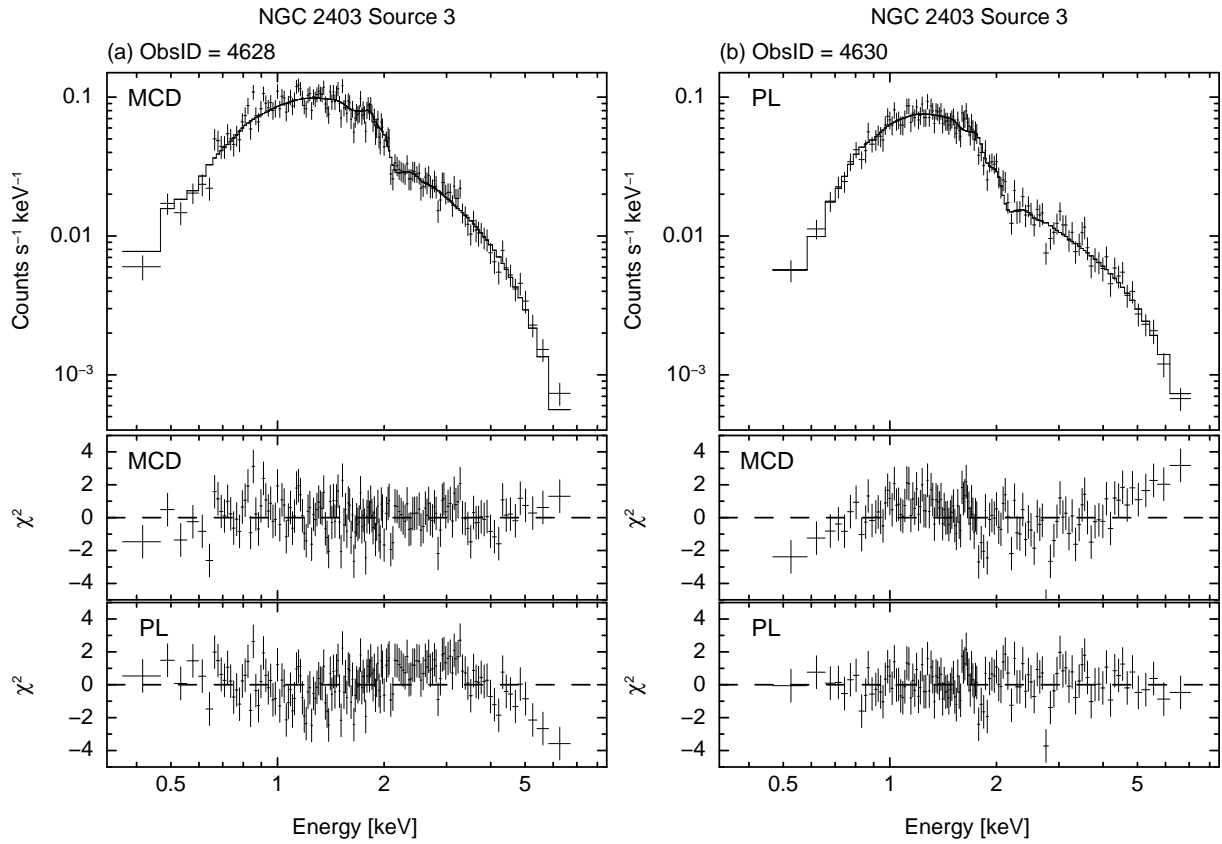

\centerline{
\FigureFile(80mm,80mm){figure5a.ps}
\FigureFile(80mm,80mm){figure5b.ps}
}
\caption{Chandra ACIS spectra of NGC 2403 Src 3,
from ObsID = 4628 (panel a) and ObsID = 4630 (panel b).
The employed model is indicated in each panel. }
\label{fig:Src3_Chandra}
\end{figure*}

\begin{figure*}[h]
\FigureFile(80mm,80mm){figure6a.ps}
\FigureFile(80mm,80mm){figure6b.ps}\\
\FigureFile(80mm,80mm){figure6c.ps}
\caption{XMM-Newton EPIC spectra of NGC 2403 Src 3, obtained on 3 occasions. 
The black, and red data points indicate the MOS and pn spectra, respectively.  
The histogram in the individual spectra indicate the best-fit MCD model. 
For the observation of ObsID = 0164560901 (panel c), 
the residuals of the variable-p disk model are also plotted.}
\label{fig:Src3_Newton}
\end{figure*}

\begin{figure}[h]
\FigureFile(80mm,80mm){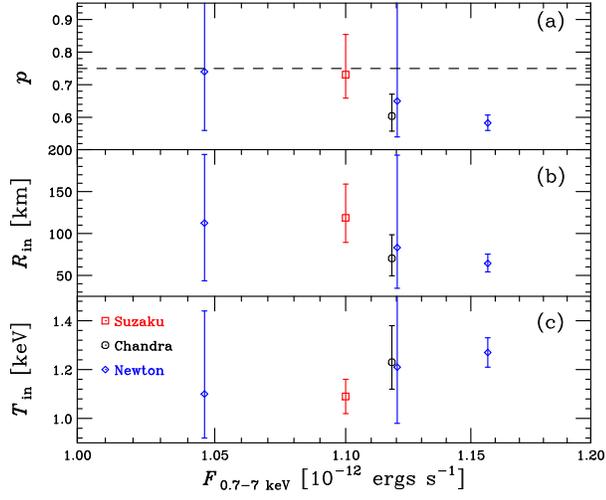}
\caption{A summary plot of the 
variable-$p$ modeling of the Suzaku, Chandra, and XMM-Newton spectra of Src 3. 
Panels (a), (b) and (c) present $p$, $R_{\rm in}$ and $T_{\rm in}$ 
as a function of the observed 0.7 -- 7 keV flux, respectively. 
The dotted line in panel (a) indicates $p = 0.75$. }
\label{fig:variable-p}
\end{figure}

\clearpage 
\appendix
\subsection{X-ray spectra of NGC~2403 Src~5} 
The X-ray signals from NGC 2403 Src 5 is 
clearly detected in figure \ref{fig:img_Suzaku}.
We breifly analyze the X-ray spectrum of Src 5,
altough we regard that a detailed study of the source
is beyond the scope of the present paper,  
due to relativley low signal statistics. 

\begin{figure}[h]
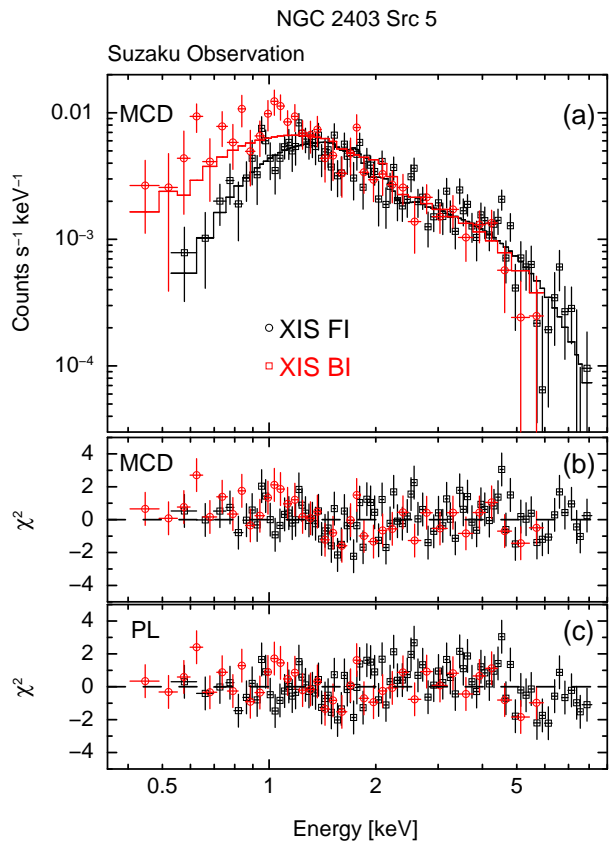

\FigureFile(80mm,80mm){figure8.ps}
\caption{Suzaku XIS spectra of NGC 2403 Src 5,
shown in the same manner as figure \ref{fig:Src3_Suzaku}.}  
\label{fig:Src5_Suzaku}
\end{figure}

The X-ray signals from Src 5 and background 
are extracted from the circles denoted as {\bf Src5} and {\bf BGD5}
in figure \ref{fig:img_Suzaku}, respectively.  
Figure \ref{fig:Src5_Suzaku} shows the XIS spectra of Src 5. 
After calculating the rmf and arf files in the same way as for Src 3,  
the MCD and PL models were applied to the spectrum.  
We did not include the thermal components from the galaxy in the fitting, 
since the signal statistics are rather low 
and Src 5~is located slightly outside the diffuse emission 
(see figure \ref{fig:img_Suzaku}). 
The absorption column density 
for the MCD model was fixed at the Galactic value, 
since it was unconstrained from the data. 

The results of this analysis are summarised in table \ref{table:Src5_Suzaku}. 
Thus, the MCD model was moderately successful, 
and yielded the disk parameters 
as $T_{\rm in} = 1.52_{-0.11}^{+0.13} $ keV and 
$R_{\rm in} = 28.6_{-3.7}^{+4.1}  ~\alpha^{1/2}$ km. 
The bolometric luminosity was estimated 
to be $L_{\rm bol} = 4.0 \times 10^{38} \alpha $ ergs s$^{-1}$. 
The PL fit was slightly less successful,
and gave a photon index of $\Gamma = 1.65_{-0.12}^{+0.13} $. 

\begin{table}[h]
\caption{Best-fit parameters for to Suzaku spectra of Src 5.}
\label{table:Src5_Suzaku}
\begin{center}
\begin{tabular}{lll}
\hline \hline 
                               & MCD                     &  PL    \\
\hline        
$N_{\rm H}$ ($10^{21}$ cm$^{-2}$) &$0.41$\footnotemark[$\dagger$] & $1.19_{-0.61}^{+0.73}$ \\        
$T_{\rm in}$ (keV)              & $1.52_{-0.11}^{+0.13}$    & --                    \\        
$R_{\rm in }$ (km)              & $28.6_{-3.7}^{+4.1} $     & --                   \\        
$\Gamma$                       & --                      & $1.65_{-0.12}^{+0.13}$ \\        
$L$ \footnotemark[$*$]
                               & $4.0$                   & $4.7$                \\        
$\chi^2/{\rm d.o.f.} $          &  $138.8/121$             & $150.5/121$           \\        
\hline        
\multicolumn{3}{@{}l@{}}{\hbox to 0pt{\parbox{60mm}{\footnotesize
\par\noindent
\footnotemark[$*$] $L_{\rm bol}$ for the MCD model, 
                  or absorption-corrected $0.5$ -- $10$ keV luminosity for the PL ones,
                  in $10^{38}$ ergs s$^{-1}$. 
\par\noindent
\footnotemark[$\dagger$] fixed at the Galactic value. 
}\hss}}
\end{tabular}
\end{center}
\end{table}

\begin{thebibliography}{}
\bibitem[Abe et al.(2005)]{4U1630‑47}
        Abe, Y.,  Fukazawa, Y., Kubota, A., 
        Kasama, D., \& Makishima, K.,
        2005, \pasj, 57, 629
\bibitem[Bamba et al.(2008)]{cross_cal2}
        Bamba A., et al. 2008, in prep.
\bibitem[Dickey \& Lockman (1990)]{NH}
        Dickey, J. M., \& Lockman, F. J.
        1990, \araa, 28, 215
\bibitem[Fabbiano \& Trinchieri(1987)]{ULX_Einstein}
        Fabbiano, G., \& Trinchieri, G., 
        1987, \apj, 315, 46
\bibitem[Feng \& Kaaret(2005)]{ULX_XMM}
        Feng, H., \& Kaaret, P.,
        2005, \apj, 633, 1052
\bibitem[Freedman \& Madore(1998)]{NGC2403_distance}
        Freedman, W. L., \& Madore, B. F., 
        1988, \apj, 332, L63 
\bibitem[Gelino \& Harrison(2003)]{GROJ0422_mass}
        Gelino, D. M., \& Harrison T. E., 
        2003, \apj, 599, 1254 
\bibitem[Hynes et al.(2003)]{GX339_mass}
	Hynes, R. I., Steeghs, D., Casares, J., 
 	Charles P. A., \& O'Brien, K.,
	2003, \apj, 583, L95
\bibitem[Ishida et al.(2008)]{cross_cal}
        Ishida M., et al. 2008 in prep. 
\bibitem[Isobe et al.(2008)]{SuzakuJ1305}
        Isobe, N., Kubota, A., Makishima, K., Gandhi, P.,
        Griffiths, R.E., Dewangan, G.C., Itho, T., \& Mizuno, T.,
        2008, \pasj, 60S, 241 
\bibitem[Ishisaki et al.(2007)]{xissimarf}
        Ishisaki, Y.,  et al. 2007, \pasj, 59, 113
\bibitem[Kokubun et al.(2007)]{HXDperform}
        Kokubun, M., et al., 2007, \pasj, 59, S53
\bibitem[Kotoku et al.(2000)]{NGC2403_ASCA}
        Kotoku, J., Mizuno, T., Kubota, A., \& Makishima, K.,
        2000, \pasj, 52, 1081
\bibitem[Koyama et al.(2007)]{XISpaper}
        Koyama K., et al., 2007, \pasj, 59, S23
\bibitem[Kubota et al.(2002)]{IC342_VHS}
        Kubota, A.,  Done, C., \& Makishima, K.
        2002, \mnras, 337, L11
\bibitem[Kubota \& Done(2004)]{XTEJ1550_VHS}
        Kubota, A., \& Done, C., 2004
        \mnras, 353, 980
\bibitem[Kubota et al.(2001)]{GROJ1655_VHS}
        Kubota, A., Makishima, K., \& Ebisawa, K.,
        2001, \apj, 560, L147  
\bibitem[Kubota \& Makishima(2004)]{XTEJ1550}
        Kubota, A., \& Makishika, K., 2004, \apj, 601, 428 
\bibitem[Kubota et al.(2001)]{IC342_transition}
        Kubota, A., Mizuno, T., Makishima, K., Fukazawa, Y.,
        Kotoku, J., Ohnishi, T., \& Tashiro, M.,
        2001, \apj, 547, L119
\bibitem[Kubota et al.(1998)]{xi}
        Kubota, A., Tanaka, Y., Makishima, K., Ueda, Y.,
        Dotani, T., Inoue, H., \& Yamaoka, Y. 1998, \pasj, 50, 667
\bibitem[Lasker et al.(1990)]{dss}
        Lasker, B. M., Sturch, C. R., McLean, B. J.; 
        Russell, J. L., Jenkner, H.; Shara, M. M.,
        \aj, 99, 2019
\bibitem[Makishima et al.(1986)]{MCD2}
        Makishima, K., Maejima, Y., Mitsuda, K., Bradt, H. V.,
        Remillard, R. A.,Tuohy, I. R., Hoshi, R., Nakagawa, M., 
        1986, \apj, 308, 635
\bibitem[Makishima et al.(2000)]{ULX_asca1}
        Makishima, K., et al., 2000, \apj, 535, 632 
\bibitem[Mineshige et al.(1994)]{pfree_disk}
        Mineshige, S., Hirano, A., Kitamoto, S., Tamada, T. T.,
        Fukue, J. 1994, \apj, 426, 308
\bibitem[Mitshuda et al.(1984)]{MCD1}
        Mitsuda, K., et al., 1984, \pasj, 36, 741
\bibitem[Mitsuda et al.(2007)]{Suzaku}
        Mitsuda, K., et al., 2007, \pasj, 59, S1
\bibitem[Miyamoto et al.(1991)]{GX339_VHS}
        Miyamoto, S., Kimura, K., Kitamoto, S., Dotani, T., \& Ebisawa, K., 
        1991, \apj, 383,784
\bibitem[Mizuno et al.(2001)]{ULX_asca2}
        Mizuno, T., Kubota, A., \& Makishima, K.,
        2001, ApJ, 554, 1282
\bibitem[Mizuno et al.(2007)]{NGC1313_Suzaku}
        Mizuno, T., et al., 2007, \pasj, 59, S257
\bibitem[Nakano et al.(2004)]{sn2004dj}
        Nakano, S., Itagaki, K., Bouma, R. J., Lehky, M., 
        \& Hornoch, K. 2004, IAU Circ., 8377
\bibitem[Ohsuga(2006)]{limit_cycle}
        Ohsuga, K., 2006, \apj, 640, 923 
\bibitem[Ohsuga et al.(2005)]{photon-trapping}
        Ohsuga, K., Mori, M., Nakamoto, T.,  \& Mineshige, S.
        2005, \apj, 628, 368
\bibitem[Pooley \& Lewin(2004)]{sn2004dj_chandra}
        Pooley, D., \& Lewin, W. H. G.,
        2004, IAU Circ., 8390, 1 
\bibitem[Read \& Ponman(2003)]{XMM_BGD}
        Read, A. M., \& Ponman, T. J.,
        2003, \aap, 409, 395
\bibitem[Reeves et al.(2007)]{MCG-5-23-16}
        Reeves, J. N., et al., 
        2007, \pasj, 59, S301
\bibitem[Schlegel \& Pannuti(2003)]{NGC2403_Chandra}
        Schlegel, E. M., \& Pannuti, T. G., 
        2003, \aj, 125, 3025
\bibitem[Serlemitsos et al.(2007)]{XRTpaper}
        Serlemitsos, P.J, et al., 2007, \pasj, 59, S9
\bibitem[Senda et al.(2008)]{NGC2403_diffuse}
        Senda, A., et al., 2008, in prep. 
\bibitem[Shahbaz et al.(1999)]{GROJ1655_mass}
	Shahbaz, T., van der Hooft, F., Casares, J.,
	Charles, P. A., \& van Paradijs, J.,
	1999, \mnras, 306, 89 
\bibitem[Shakura \& Sunyaev(1973)]{standard_disk}
        Shakura N. I., \& Sunyaev, R. A. 1973, \aap, 24, 337
\bibitem[Shimura \& Takahara(1995)]{kappa}
        Shimura, T., \& Takahara, F.
        1995, \apj, 445, 780
\bibitem[Sivakoff et al.(2008)]{ULX_in_CenA}
        Sivakoff, G. R., et al. 2008, \apj, 677, L27
\bibitem[Stobbart et al.(2006)]{ULX_XMM1}
        Stobbart, A. -M., Roberts, T. P., \& Wilms, J.,
        2006, \mnras, 368, 397
\bibitem[Takahashi et al.(2007)]{HXDdesign}
        Takahashi, T., et al.,  2007, \pasj, 59, S35
\bibitem[Tanaka(1997)]{Index_in_low/hard}
        Tanaka, Y., 1997, in Accretion Disks -- New Aspects, 
        Proceedings of the EARA Workshop Held in Garching, Germany, 
        (Lecture Notes in Phys., 487; Springer Berlin / Heidelberg) 
\bibitem[Titarchuk(1994)]{comptt}
	Titarchuk, L., 1994, \apj, 434, 570
\bibitem[Tsunoda et al.(2006)]{M81X9}
        Tsunoda, N., Kubota, A., Namiki, M., Sugiho, M.,
        Kawabata, K., \& Kazuo, M.
        2006, \pasj, 58, 1081
\bibitem[Vierdayanti et al.(2008)]{mass_in_slim}
        Vierdayanti, K., Watarai, K., \& Mineshige, S.,
        2008, \pasj, 60, 653
\bibitem[Watarai et al.(2000)]{slimdisk}
        Watarai, K., Fukue, J., \& Mineshige, S.
        2000, \pasj, 52, 133
\bibitem[Winter et al.(2006)]{ULX_XMM2}
        Winter, L. M., Mushotzky, R. F., \& Reynolds, C .S., 
        2006, \apj, 649, 730
\end{thebibliography}
\end{document}